\RequirePackage{ifpdf}
\ifpdf 
\documentclass[pdftex]{sigma}
\else
\documentclass{sigma}
\fi

\numberwithin{equation}{section}

\newcommand{\eqdef}{\stackrel{\rm def}{=}}

\begin{document}

\allowdisplaybreaks

\renewcommand{\thefootnote}{$\star$}

\renewcommand{\PaperNumber}{104}

\FirstPageHeading

\ShortArticleName{Bethe Ansatz Solutions to Quasi Exactly Solvable
Dif\/ference Equations}

\ArticleName{Bethe Ansatz Solutions to Quasi Exactly Solvable\\
Dif\/ference Equations\footnote{This paper is a
contribution to the Proceedings of the 5-th Microconference
``Analytic and Algebraic Me\-thods~V''. The full collection is
available at
\href{http://www.emis.de/journals/SIGMA/Prague2009.html}{http://www.emis.de/journals/SIGMA/Prague2009.html}}}

\Author{Ryu SASAKI~$^\dag$, Wen-Li YANG~$^{\ddag\S}$ and Yao-Zhong ZHANG~$^\S$}

\AuthorNameForHeading{R.~Sasaki, W.-L.~Yang and Y.-Z.~Zhang}

\Address{$^\dag$~Yukawa Institute for Theoretical Physics, Kyoto University, Kyoto 606-8502, Japan}
\EmailD{\href{mailto:ryu@yukawa.kyoto-u.ac.jp}{ryu@yukawa.kyoto-u.ac.jp}}

\Address{$^\ddag$~Institute of Modern Physics, Northwest University, Xian 710069, P.R.~China}
\EmailD{\href{mailto:wlyang@nwu.edu.cn}{wlyang@nwu.edu.cn}}

\Address{$^\S$~School of Mathematics and Physics, The University of Queensland,\\
\hphantom{$^\S$}~Brisbane, QLD 4072, Australia}
\EmailD{\href{mailto:yzz@maths.uq.edu.au}{yzz@maths.uq.edu.au}}

\ArticleDates{Received September 20, 2009, in f\/inal form November 10, 2009;  Published online November 18, 2009}

\Abstract{Bethe ansatz formulation is presented for several explicit
examples of quasi exactly solvable dif\/ference equations of one
degree of freedom which  are introduced recently by one of the
present authors. These equations are deformation of the well-known
exactly solvable dif\/ference equations of the Meixner--Pollaczek,
continuous Hahn, continuous dual Hahn, Wilson and Askey--Wilson
polynomials. Up to an overall factor of the so-called pseudo
ground state wavefunction, the eigenfunctions within the exactly
solvable subspace are  given by
  polynomials whose roots are solutions of the associated Bethe
ansatz equations. The corresponding eigenvalues are expressed in
terms of these roots.}


\Classification{35Q40; 37N20; 39A70; 82B23}

\section{Introduction}
\label{intro}

Bethe ansatz method is one of the well-known solution methods for
exactly solvable quantum systems (see, e.g.~\cite{susyqm})  as well as for
various spin models and statistical lattice models. In recent
years the concept of exactly solvable quantum systems was
drastically enlarged to include various examples of the so-called
`discrete' quantum mechanical systems \cite{os4,os5,os7,os9,os13,os12},
 in which the
Schr\"odinger equation is a dif\/ference equation instead of
dif\/ferential. Known examples of exactly solvable `discrete'
quantum mechanics are deformations of exactly solvable quantum
mechanics, in which the momentum operators appear in exponentiated
forms instead of polynomials in ordinary quantum mechanics. Their
eigenfunctions are ($q$-)Askey scheme of hypergeometric orthogonal
polynomials \cite{askey,koeswart}, which are deformations of the
classical orthogonal polynomials, e.g.\ the Hermite, Laguerre
and Jacobi polynomials. It is interesting to note that these
examples are exactly solvable both in the Schr\"odinger and
Heisenberg pictures \cite{os7,os9}. That is, for each Hamiltonian
of these examples, the Heisenberg operator solution for a special
coordinate called the `{\em sinusoidal}' coordinate can be
constructed, as well as the complete set of the eigenvalues and
the corresponding eigenfunctions. The eigenfunctions consist of
the above-mentioned \mbox{($q$-)Askey} scheme of hypergeometric
orthogonal polynomials in the  `sinusoidal' coordinate.
See \cite{os13,os12} for a~comprehensive introduction of the `discrete' quantum mechanics and the recent developments.
The domain of the Hamiltonian, its hermiticity (self-adjointness),
the exact Heisenberg ope\-ra\-tor solutions, the creation-annihilation operators and the dynamical symmetry algebras are explained in some detail.

In this paper we present Bethe ansatz formulations and solutions for a family of
Quasi-Exactly Solvable (QES) dif\/ference equations,  which were recently introduced by
one of the present authors \cite{deltaqes,newqes}.
One of the main purposes of the present paper is to provide a good list of explicit
Bethe-ansatz equations for the quasi-exactly solvable dif\/ference equations
\eqref{BAE-1},  \eqref{BAE-2-1},  \eqref{BAE-2-2}, \eqref{BAE-3},
\eqref{BAE-3-2}, and \eqref{BAE-4}. The list will be helpful for future research in
various contexts of mathematical physics.
A quantum
mechanical system is called quasi-exactly sol\-vable (QES), if only
a f\/inite number ($\ge2$) of eigenvalues and corresponding
eigenvectors can be obtained exactly \cite{Ush,turb}. Among
various characterisation/identif\/ication of QES systems
\cite{Ush,turb,N-SUSY,st1}, we promoted a simple view that a QES
system is obtained by a certain deformation of an exactly solvable
quantum system. As shown explicitly for ordinary quantum mechanics
by Sasaki--Takasaki~\cite{st1}, the deformation procedure applies
to systems of many degrees of freedom as well as for single degree
of freedom systems. This is in sharp contrast to the $sl(2,R)$
characterisation~\cite{turb}, whose applicability is limited to
essentially single degree of freedom systems. Recently the
deformation procedure was applied to exactly solvable `discrete'
quantum systems of one and many degrees of freedom to obtain
corresponding `discrete' QES systems~\cite{deltaqes,newqes,os10}.
In this paper we present Bethe ansatz solutions for these QES dif\/ference equations.

There are only a very limited number of examples of (quasi)
exactly solvable dif\/ference equations which have been solved so
far by the Bethe ansatz method. To our knowledge, the Bethe ansatz
solutions were known only for some QES dif\/ference equations in
connection with $U_q(sl(2))$ \cite{Wiegmann}, and for exactly
solvable dif\/ference equations \cite{Fel96,Hou03, Man07} of the
elliptic Rusijsenaars--Schneider system. Here we apply the Bethe
ansatz formulation to  several explicit examples of QES dif\/ference
equations \cite{deltaqes,newqes} as deformation of {\em exactly
solvable} `discrete' quantum mechanics~\cite{os4, os5}, which
are dif\/ference analogues of the well-known quasi exactly solvable
systems, the harmonic oscillator (with/without the centrifugal
potential) deformed by a sextic potential and the $1/\sin^2x$
potential deformed by a $\cos2x$ potential. As will be shown
explicitly in the main text, these Bethe ansatz equations
\eqref{BAE-1},  \eqref{BAE-2-1},  \eqref{BAE-2-2}, \eqref{BAE-3},
\eqref{BAE-3-2}, and~\eqref{BAE-4} can be considered as
deformations of the equations \eqref{rBAE-MP}, \eqref{rBAE-3},
\eqref{rBAE-AW} determining the roots of the corresponding
($q$-)Askey scheme of hypergeometric orthogonal polynomials
\cite{koeswart} (the Meixner--Pollaczek, continuous Hahn,
continuous dual Hahn, Wilson and Askey--Wilson polynomials and
their restrictions)  which constitute the eigenfunctions of the
undeformed exactly solvable quantum systems.
The general structure of the quasi exactly solvable dif\/ference equations to be discussed in this paper  and their solutions were explained in some detail
in \cite{deltaqes,newqes,os10}, inclu\-ding the domains of the Hamiltonians, the
Hilbert spaces and hermiticity and the role played by the pseudo ground state wavefunctions.

\looseness=-1
This paper is organised as follows. In Section~\ref{Meix-Poll}, a QES discrete
quantum mechanics is solved in the Bethe ansatz formalism. The
Hamiltonian of the system is obtained by `crossing' those of the
Meixner--Pollaczek and the continuous Hahn polynomials as derived in
\cite{newqes}. Section~\ref{difex1} provides the Bethe ansatz formulation
of the QES discrete quantum mechanical systems which are
deformations of the harmonic oscillator with a sextic potential
as derived in~\cite{deltaqes}.
The corresponding eigenfunctions are deformations of the
Meixner--Pollaczek and the continuous Hahn polynomials. Section~\ref{difsex}
gives the Bethe ansatz solutions of the QES discrete quantum
mechanical systems which are deformations of the harmonic
oscillator with a centrifugal barrier and a sextic potential as derived in \cite{deltaqes}. The
corresponding eigenfunctions are deformations of  the continuous
dual Hahn and the Wilson polynomials. Section~\ref{diftrig} of\/fers a Bethe
ansatz solution to a dif\/ference equation analogue of a QES system
with the $1/\sin^2x$ potential deformed by a~$\cos2x$ potential as derived in~\cite{deltaqes}.
The corresponding eigenfunctions are deformations of  the
Askey--Wilson polynomials and their various restrictions~\cite{askey,koeswart}. The f\/inal section is for a~summary and
comments.


\section[Difference equation of the Meixner-Pollaczek type]{Dif\/ference equation of the Meixner--Pollaczek type}
  \label{Meix-Poll}

In this section we will discuss Bethe ansatz solutions for the
discrete quantum mechanics obtained by  deforming that of  the
Meixner--Pollaczek polynomials in \cite{newqes}. To be more
precise, the corresponding discrete quantum mechanics is obtained
by  crossing those of the Meixner--Pollaczek and the continuous
Hahn polynomials, that is, with the quadratic potential function
of the continuous Hahn polynomial multiplied by a constant phase
factor $e^{-i\beta}$ of the Meixner--Pollaczek type. It was shown
in \cite{newqes} that this system is quasi exactly solvable. The
corresponding Hamiltonian is:
\begin{gather}
\mathcal{H} \eqdef\sqrt{V(x)}\,e^{-i\partial_x}\sqrt{V(x)^*}
   +\sqrt{V(x)^*}\,e^{+i\partial_x}\sqrt{V(x)}-V(x)-V(x)^*+\alpha_{\mathcal M}x
   \label{firstHam}\\
    \phantom{\mathcal{H}}{} \, =A^\dagger{A}+\alpha_{\mathcal M}x,\qquad \alpha_{\mathcal M}
   \eqdef -2\mathcal{M}\sin\beta,\qquad \mathcal{M}\in\mathbb{Z}_+,
   \label{factformadded}\\
    A^{\dagger}
    \eqdef\sqrt{V(x)}\,e^{-\frac{i}{2}\partial_x}
   -\sqrt{V(x)^*}\,e^{\frac{i}{2}\partial_x},\quad
  A \eqdef e^{-\frac{i}{2}\partial_x}\sqrt{V(x)^*}
   -e^{\frac{i}{2}\partial_x}\sqrt{V(x)},
   \label{defAAdag}\\
\phantom{A^{\dagger}     \eqdef}{} \ V(x) \eqdef(a_1+ix)(a_2+ix)e^{-i\beta}, \qquad
a_1,a_2\in\mathbb{C},
  \quad  \mbox{Re}(a_1)>0, \quad \mbox{Re}(a_2)>0.\nonumber
\end{gather}
It should be noted that the Hamiltonian is no longer positive
semi-def\/inite but the hermiticity is preserved. Let us introduce
the so-called {\em pseudo ground state} wavefunction $\phi_0(x)$
\cite{deltaqes,newqes}:
\begin{gather*}
   \phi_0(x)\eqdef
   e^{\beta
   x}\sqrt{\Gamma(a_1+ix)\Gamma(a_2+ix)\Gamma(a_1^*-ix)\Gamma(a_2^*-ix)},
\end{gather*}
as the zero mode of the $A$ operator~\eqref{defAAdag},
$A\phi_0=0$. The similarity transformed Hamiltonian
$\widetilde{\mathcal H}$ in terms of $\phi_0$,
\begin{gather}
  \widetilde{\mathcal{H}} \eqdef
   \phi_0^{-1}\circ\mathcal{H}\circ\phi_0
=V(x)\big(e^{-i\partial_x}-1\big)+V(x)^*\big(e^{i\partial_x}-1\big)+\alpha_{\mathcal
M}x
   \label{tilH3}\\
\phantom{\widetilde{\mathcal{H}}}{} \, =  (a_1+ix)(a_2+ix)e^{-i\beta}\big(e^{-i\partial_x}\!-1\big)
  +(a_1^*-ix)(a_2^*-ix)e^{i\beta}\big(e^{i\partial_x}\!-1\big)
  -2\mathcal{M}\sin\beta x,\nonumber
\end{gather}
acts on the polynomial part of the wavefunction. In the exactly
solvable Meixner--Pollaczek case with $V(x)=(a+ix)e^{-i\beta}$ and
the continuous Hahn case with $V(x)=(a_1+ix)(a_2+ix)$, the
eigenfunctions are of the form
\begin{gather*}
\phi_0(x)P(\eta(x))
\end{gather*}
in which $P(\eta(x))$ is a polynomial in
\[
\eta(x)=x.
\]
After the deformation, it is obvious that $\widetilde{\mathcal H}$
maps a polynomial in $\eta(x)=x$  into another and it is easy to
verify
\begin{gather*}
  \widetilde{\mathcal{H}} x^n=2(-\mathcal{M}+n)\sin\beta\,
x^{n+1}+\mbox{lower order terms}, \qquad n\in\mathbb{Z}_+.
\end{gather*}
This means that the system is not exactly solvable without the
compensation term, but it is quasi exactly solvable, since
$\widetilde{\mathcal H}$ has an invariant polynomial subspace of
degree $\mathcal{M}$:
\begin{gather}
  \widetilde{\mathcal{H}} {\mathcal V}_{\mathcal M} \subseteq {\mathcal
   V}_{\mathcal M},\label{subspc1}\\
  {\mathcal V}_{\mathcal M}
 \eqdef  \mbox{Span}\left[1, x, x^2,\ldots,x^{\mathcal
     M}\right],\qquad \mbox{dim}{\mathcal V}_{\mathcal M}={\mathcal M}+1.
     \label{Vdimform1}
\end{gather}
Let $\Psi(x)$ be one of the eigenfunctions of
$\widetilde{\mathcal{H}}$ and $E$ be the corresponding eigenvalue:
\begin{gather*}
  \widetilde{\mathcal{H}} \Psi(x)=E \Psi(x),
\end{gather*}
namely,
\begin{gather}
  (a_1^*-ix)(a_2^*-ix)e^{i\beta}\left(\Psi(x+i)-\Psi(x)\right)
  \nonumber\\
  \qquad{}+(a_1+ix)(a_2+ix)e^{-i\beta}\left(\Psi(x-i)
 - \Psi(x)\right)
    -
   2{\mathcal M}\sin\beta x\Psi(x)=E \Psi(x).
  \label{eigen-ned}
\end{gather}
Equations (\ref{subspc1}) and (\ref{Vdimform1}) imply that the
eigenfunctions in the subspace ${\mathcal V}_{\mathcal M}$ have
the following form
\begin{gather}
  \Psi(x)=\prod_{l=1}^{{\mathcal
M}}\left(\eta(x)-\eta(x_l)\right)=\prod_{l=1}^{{\mathcal
M}}(x-x_l),\label{Eign-1}
\end{gather}
where $\left\{x_l\,|\,l=1,\ldots,{\mathcal M}\right\}$ are some
parameters which will be specif\/ied later by the associated Bethe
ansatz equations (\ref{BAE-1}) below. Substituting the above
equation into (\ref{eigen-ned}) and dividing both sides by
$\Psi(x)$, we have
\begin{gather}
  E=-V(x)-V(x)^*-2{\mathcal M}\sin\beta\,x+V(x)
   \prod_{l=1}^{{\mathcal M}}\frac{x-x_l-i}{x-x_l}
   +V(x)^*\prod_{l=1}^{{\mathcal
   M}}\frac{x-x_l+i}{x-x_l}.\label{E-1}
\end{gather}
The r.h.s.\ of (\ref{E-1}) is a meromorphic function of $x$,
whereas the l.h.s.\ is a constant. To make them equal, we must null
the residues of the r.h.s.\ It is easy to see that the singularities of
the r.h.s.\ only appear at $x=x_j$, $j=1,\ldots, {\mathcal M}$ and
$x=\infty$. The residues at $x=x_j$ vanish if the parameters
$\left\{x_j\right\}$ satisfy the following Bethe ansatz equations
\begin{gather}
  \prod_{l\neq j}^{{\mathcal M}}\frac{x_j-x_l-i}{x_j-x_l+i}
 =\frac{V(x_j)^*}{V(x_j)}\,\frac{\eta(x_j+i)-\eta(x_j)}{\eta(x_j)-\eta(x_j-i)}
   \label{BAgenform1}\\
\hphantom{\prod_{l\neq j}^{{\mathcal M}}\frac{x_j-x_l-i}{x_j-x_l+i}}{}
=\frac{V(x_j)^*}{V(x_j)}
  e^{2i\beta}\frac{(a_1^*-ix_j)(a_2^*-ix_j)}{(a_1+ix_j)(a_2+ix_j)},
  \qquad j=1,\ldots,{\mathcal M}.\label{BAE-1}
\end{gather}
Throughout this paper  we use the complex conjugate potential
function $V(x)^*$ in the `analytical' sense in $x$, that is, for
complex $x$
\[
  V(x)=(a_1+ix)(a_2+ix)e^{-i\beta},\qquad
V(x)^*=(a_1^*-ix)(a_2^*-ix)e^{i\beta},\qquad x\in\mathbb{C}.
\]
This convention is necessary for the above two equations
\eqref{BAgenform1}, \eqref{BAE-1} to be valid, since the Bethe
roots $\{x_j\}$ are in general complex. One can check that the
r.h.s.\ of~\eqref{E-1} is indeed regular at $x=\infty$, i.e., the
residue at $x=\infty$ vanishes. By the Liouville theorem the
r.h.s.\ of~(\ref{E-1}) is a~constant provided that (\ref{BAE-1}) is
satisf\/ied. One can  get the value of the corresponding eigenvalue~$E$ by taking the limit of $x\rightarrow \infty$ for the r.h.s.\ of~(\ref{E-1}). Here we present the result:
\begin{gather}
  E={\mathcal M}({\mathcal M}-1)\cos\beta+{\mathcal M}
  \big((a_1+a_2)e^{-i\beta}+(a_1^*+a_2^*)e^{i\beta}\big)
  +2\sin\beta\,\sum_{l=1}^{{\mathcal M}}x_l,
  \label{Eigenvalue-1}
\end{gather}
where $\left\{x_l\right\}$ satisfy the Bethe ansatz equation
(\ref{BAE-1}). The f\/inal term can be written as
\begin{equation}
2\sin\beta \sum_{l=1}^{{\mathcal M}}\eta(x_l). \label{symmetric1}
\end{equation}
The wavefunction  $\Psi(x)$ (\ref{Eign-1}) becomes the
eigenfunction of $\widetilde{\mathcal{H}}$ in the subspace
${\mathcal V}_{\mathcal M}$ (\ref{Vdimform1}) provided that the
roots of the polynomial $\Psi(x)$ (\ref{Eign-1}) are the solutions
of (\ref{BAE-1}), and then the corresponding eigenvalue is given
by (\ref{Eigenvalue-1}). Since all the roots $\{x_l\}$ are on the
same footing, it is natural that the eigenvalue $E$ depends on the
symmetric combination of them \eqref{symmetric1}.

A few corollaries ensue from these results. For the special case
of $\beta=0$, the Hamiltonian~\eqref{firstHam} is exactly solvable
and the corresponding eigenvectors are related to  the continuous
Hahn polynomials. In fact, we obtain from~\eqref{Eigenvalue-1}
\begin{equation}
\lim_{\beta\to0}E=\mathcal{M}(\mathcal{M}+a_1+a_2+a_1^*+a_2^*-1),
\label{eformCH}
\end{equation}
  which is the eigenvalue corresponding to the degree  $\mathcal{M}$
continuous Hahn polynomial \cite{os4,os5,os13}. The Bethe ansatz
equation  \eqref{BAE-1} now determines the zeros of the continuous
Hahn polynomial. For $a_1,a_2\in\mathbb{R}_+$, and $a_2\to \infty$
limit, the above Hamiltonian
  $\mathcal{H}$ \eqref{firstHam} divided by $a_2$ reduces to an
exactly solvable one corresponding to the Meixner--Pollaczek
polynomials \cite{os4,os5,os13}.
  Correspon\-dingly the eigenvalue formula \eqref{Eigenvalue-1} gives
\begin{equation}
  \lim_{a_2\to\infty} E/a_2=2\mathcal{M}\cos\beta.
  \label{eformMP}
\end{equation}
  The latter is the eigenvalue of the  degree $\mathcal{M}$
Meixner--Pollaczek polynomial \cite{os13}. Note that the parameter
$\beta$ is related to the standard parameter $\phi$ of the
  Meixner--Pollaczek polynomials as $\beta=\frac{\pi}{2}-\phi$. The
corresponding Bethe ansatz equation
\begin{gather}
  \prod_{l\neq j}^{{\mathcal M}}\frac{x_j-x_l-i}{x_j-x_l+i}
  =e^{2i\beta}\frac{(a_1-ix_j)}{(a_1+ix_j)},
  \qquad j=1,\ldots,{\mathcal M},
  \label{rBAE-MP}
\end{gather}
now determines the zeros of the degree $\mathcal{M}$
Meixner--Pollaczek polynomial. As shown in \S~4 of~\cite{os4}, this
equation reduces to that determines the zeros of the Hermite
polynomial in an appropriate limit.

\section{Dif\/ference equation analogue of harmonic oscillator\\
deformed by sextic potential} \label{difex1}

There are two types of dif\/ference equations which are dif\/ference
analogues of the sextic potential Hamiltonian~\cite{deltaqes}.  The
Hamiltonian is given by
\begin{gather}
   \mathcal{H} \eqdef \sqrt{V(x)}\,e^{-i\partial_x}\sqrt{V(x)^*}
   + \sqrt{V(x)^*}\,e^{i\partial_x}\sqrt{V(x)}
   -(V(x)+\!V(x)^*) + \alpha_\mathcal{M}(x),
   \label{H1}\\
\text{Type I}:\quad   V(x)\eqdef(a+i x)(b+i x)V_0(x),\qquad
  V_0(x)\eqdef c+ ix,\qquad a,b,c\in\mathbb{R}_+,
  \label{typeIV}\\
\phantom{\text{Type I}:\quad}{} \  \alpha_{\mathcal M}(x)\eqdef 2 \mathcal{M} x^2,\nonumber\\
\text{Type II}:\quad    V(x)\eqdef(a+i x)(b+i x)V_0(x),\nonumber\\
\phantom{\text{Type II}:\quad}{} \ V_0(x)\eqdef(c+ ix)(d+i x),\qquad a,b,c,d\in\mathbb{R}_+,
\label{typeIIV}\\
\phantom{\text{Type II}:\quad}{} \  \alpha_\mathcal{M}(x)\eqdef
\mathcal{M}\left(\mathcal{M}-1+2(a+b+c+d)\right)x^2.
\label{typeIIVa}
\end{gather}
Here as usual $V(x)^*$ is the `analytical' complex conjugate of~$V(x)$.

\subsection{Type I theory}
Here we will consider the dif\/ference equation of type~I, while the
  dif\/ference equation of type~II will be given in the next subsection.
If $V$ is replaced by $V_0$ in~\eqref{typeIV} and the last term in~(\ref{H1}),~$\alpha_\mathcal{M}(x)$, is removed, $\mathcal{H}$
becomes the exactly solvable Hamiltonian of a dif\/ference analogue
of the harmonic oscillator, or the {\em deformed harmonic
oscillator} in `discrete' quantum mechanics~\cite{os4,os5}. Its
eigenfunctions consist of the Meixner--Pollaczek polynomials, with
a special phase angle $\beta=0$, which is a deformation of the
Hermite polynomials~\cite{os4,os5,degruij}. The quadratic
polynomial factor $(a+i x)(b+i x)$ can be considered as
multiplicative deformation, although the parame\-ters~$a$,~$b$ and~$c$ are on the equal footing. On the other hand one can consider
it as a multiplicative deformation by a linear polynomial in~$x$:
\[
V(x)=(a+i x)V_{01}(x),\qquad V_{01}(x)\eqdef(b+i x)(c+i x),
\]
with $V_{01}$ describing another dif\/ference version of an exactly
solvable  analogue of the harmonic oscillator \cite{os4,os5}. Its
eigenfunctions consist of the continuous Hahn polynomials.

Next let us introduce the similarity transformation in terms of
the  pseudo ground state wavefunction $\phi_0(x)$ as the zero mode
of the $A$ operator \eqref{defAAdag}, $A\phi_0=0$:
\begin{gather}
  \phi_0(x) \eqdef \sqrt{\Gamma(a+i x)\Gamma(a-i x)
  \Gamma(b+i x)\Gamma(b-i x)\Gamma(c+i
   x)\Gamma(c-i x)},\nonumber\\
  \widetilde{\mathcal{H}} \eqdef
   \phi_0^{-1}\circ\mathcal{H}\circ\phi_0
   =V(x)\big(e^{-i\partial_x}-1\big)+V(x)^*\big(e^{i\partial_x}-1\big)
   +2\mathcal{M} x^2.
   \label{tilH}
\end{gather}
Since the parity is conserved, that is
\begin{gather*}
\left.\mathcal{H}\right|_{x\to -x}=\mathcal{H},
\end{gather*}
it is easy to verify the action of the Hamiltonian
$\widetilde{\mathcal{H}}$ (\ref{tilH})  on monomials of~$x$:
\begin{gather}
  \widetilde{\mathcal{H}} x^n=\left\{\begin{array}{lll}
 \displaystyle  \sum_{j=0}^{[n/2+1]}a_{n, j}x^{n+2-2j},& n\leq \mathcal{M}-2,&
  a_{n,j}\in\mathbb{R},\vspace{2mm}\\
 \displaystyle   \sum_{j=0}^{[\mathcal{M}/2]}a'_{n,\,j}x^{\mathcal{M}-2j},&  n=
  \mathcal{M},&a'_{n,j}\in\mathbb{R}.
  \end{array}\right.
   \label{Hactx}
\end{gather}
Here $[m]$ is the standard Gauss' symbol denoting the greatest
integer  not exceeding or equal to~$m$. According to  the parity
of the polynomials, there are two types of
  invariant subspaces~$ {\mathcal V}_{\mathcal M}$ of
$\widetilde{\mathcal{H}}$:
\begin{gather}
   \widetilde{\mathcal{H}}\,{\mathcal V}_{\mathcal
M}\subseteq
  {\mathcal V}_{\mathcal M},
  \label{subspc2}\\
     {\mathcal V}_{\mathcal M}
   \eqdef
  \left\{\begin{array}{ll}
  \mbox{Span}\left[1,\eta(x),\ldots,\eta(x)^{k},\ldots,\eta(x)^{\mathcal
     M/2}\right],& \mathcal{M}:\mbox{even},\vspace{2mm}\\
x\;
\mbox{Span}\left[1,\eta(x),\ldots,\eta(x)^{k},\ldots,\eta(x)^{({\mathcal
     M}-1)/2}\right],& \mathcal{M}:\mbox{odd},
   \end{array}\right.\qquad \eta(x)=x^2,
     \label{eq:V0def2}\\
 \mbox{dim}{\mathcal V}_{\mathcal M}=\left\{
   \begin{array}{ll}
    {\mathcal M}/2+1,& {\mathcal M}: \mbox{even},\\
    ({\mathcal M}+1)/2,& {\mathcal M}: \mbox{odd}.
   \end{array}
    \right.\nonumber
\end{gather}

\subsubsection[The case of even ${\mathcal M}$]{The case of even $\boldsymbol{{\mathcal M}}$}

Let us introduce a positive integer $N$ such that ${\mathcal
M}=2N$. Equations (\ref{subspc2}) and (\ref{eq:V0def2}) imply that
the eigenfunctions of $\widetilde{\mathcal{H}}$ in the subspace
${\mathcal V}_{\mathcal M}$ are of the form
\begin{gather}
  \Psi(x)=\prod_{l=1}^{N}\left(\eta(x)-\eta(x_l)\right)=\prod_{l=1}^{N}(x-x_l)(x+x_l).\label{Eign-2-1}
\end{gather}
Analogous calculation shows that the polynomial $\Psi(x)$ becomes
the eigenfunction of $\widetilde{\mathcal{H}}$ if the roots of the
polynomial  satisfy the Bethe ansatz equations
\begin{gather}
  \prod_{l\neq j}^{N}
    \frac{(x_j-x_l-i)(x_j+x_l-i)}{(x_j-x_l+i)(x_j+x_l+i)}
 =\frac{V(x_j)^*}{V(x_j)}\,\frac{\eta(x_j+i)-\eta(x_j)}{\eta(x_j)-\eta(x_j-i)}
   \label{BAgenform2}\\
\qquad{}
=\frac{(a-ix_j)(b-ix_j)(c-ix_j)(2x_j+i)}
      {(a+ix_j)(b+ix_j)(c+ix_j)(2x_j-i)},\qquad j=1,\ldots, N.
      \label{BAE-2-1}
  \end{gather}
The corresponding eigenvalue $E$ is given by
\begin{gather}
   E = \frac{1}{3}{\mathcal M}({\mathcal M}-1)({\mathcal M}-2)
      +(a+b+c){\mathcal M}({\mathcal M}-1)  +2(ab+ac+bc){\mathcal M}
     -4\sum_{l=1}^Nx^2_l,
     \label{Eigenvalue-2-1}
  \end{gather}
where $\{x_l\}$ satisfy the Bethe ansatz equations
(\ref{BAE-2-1}). Again the f\/inal term is symmetric in $\{x_j\}$
and can be written as $-4\sum_{j=1}^N\eta(x_j)$.

\subsubsection[The case of odd ${\mathcal M}$]{The case of odd $\boldsymbol{{\mathcal M}}$}

Let us introduce a positive integer $N$ such that ${\mathcal
M}=2N+1$. (\ref{subspc2}) and (\ref{eq:V0def2}) imply that the
eigenfunctions of $\widetilde{\mathcal{H}}$ in the subspace
${\mathcal V}_{\mathcal M}$ are of the form
\begin{gather}
  \Psi(x)=x \prod_{l=1}^{N}\left(\eta(x)-\eta(x_l)\right)=x \prod_{l=1}^{N}(x-x_l)(x+x_l).\label{Eign-2-2}
\end{gather}
Analogous calculation shows that the polynomial $\Psi(x)$ becomes
the eigenfunction of $\widetilde{\mathcal{H}}$ if the roots of the
polynomial  satisfy the Bethe ansatz equations
\begin{gather}
\frac{(x_j-i)}{(x_j+i)} \prod_{l\neq j}^{N}
    \frac{(x_j-x_l-i)(x_j+x_l-i)}{(x_j-x_l+i)(x_j+x_l+i)}
 =\frac{V(x_j)^*}{V(x_j)}\,\frac{(\eta(x_j+i)-\eta(x_j))}{(\eta(x_j)-\eta(x_j-i))}
   \label{BAgenform3}\\
\qquad{} =\frac{(a-ix_j)(b-ix_j)(c-ix_j)(2x_j+i)}
      {(a+ix_j)(b+ix_j)(c+ix_j)(2x_j-i)},\qquad j=1,\ldots, N.\label{BAE-2-2}
  \end{gather}
  The corresponding eigenvalue $E$ is given by
\begin{gather}
   E = \frac{1}{3}{\mathcal M}({\mathcal M}-1)({\mathcal M}-2)
      +(a+b+c){\mathcal M}({\mathcal M}-1)
 +2(ab+ac+bc){\mathcal M}
     -4\sum_{l=1}^Nx^2_l,\label{Eigenvalue-2-2}
\end{gather}
where $\{x_l\}$ satisfy the Bethe ansatz equations
(\ref{BAE-2-2}). The expression for the eigenvalue is exactly the
same as the even $\mathcal M$ case \eqref{Eigenvalue-2-1}.

In this example, exactly solvable limits are also obtained by
making one or two parameters go to inf\/inity; for example
$a\to\infty$ or both $a\to\infty$ and $b\to\infty$. In the former
case ($a\to\infty$), the scaled Hamiltonian \eqref{H1}
$\mathcal{H}/a$ gives that of the continuous Hahn polynomials with
real parameters $b$ and $c$. The eigenvalue formulas
\eqref{Eigenvalue-2-1} and  \eqref{Eigenvalue-2-2} reduce to that
of the continuous Hahn polynomials \cite{os4,os5,os13}:
\begin{equation}
  \lim_{a\to\infty} E/a=\mathcal{M}(\mathcal{M}+2(b+c)-1).
  \label{eformcH2}
\end{equation}
The Bethe ansatz equations \eqref{BAE-2-1} and \eqref{BAE-2-2} are
equivalent to \eqref{BAE-1} with $\beta=0$ and
$a_1=b\in\mathbb{R}_+$, $a_2=c\in\mathbb{R}_+$. This assertion can
be easily verif\/ied since the solutions of  \eqref{BAE-1} are
always paired $\{x_j,-x_j\}$ including a zero  $x_j=0$ for odd
$\mathcal{M}$. In the latter case ($a\to\infty$, $b\to\infty$),
the scaled Hamiltonian \eqref{H1} $\mathcal{H}/(ab)$ gives that of
the Meixner--Pollaczek polynomials with $\beta=0$, $a=c$. The
eigenvalue formulas \eqref{Eigenvalue-2-1} and
\eqref{Eigenvalue-2-2} reduce to that of the  Meixner--Pollaczek
polynomials \cite{os4,os5,os13}:
\begin{equation}
  \lim_{a,b\to\infty} E/(ab)=2\mathcal{M}.
  \label{eformMP2}
\end{equation}
Again the Bethe ansatz equations \eqref{BAE-2-1} and
\eqref{BAE-2-2} are equivalent to \eqref{BAE-1} with $\beta=0$ and
$a_1=c\in\mathbb{R}_+$.

\subsection{Type II theory}
  \label{difex2}
Another dif\/ference analogue of the sextic potential Hamiltonian
has the same form as (\ref{H1}), with only the potential function
$V(x)$ and the compensation term $\alpha_\mathcal{M}(x)$ are
dif\/ferent:
\begin{gather*}
  V(x) \eqdef (a+i x)(b+i x)V_0(x),\qquad
V_0(x)\eqdef(c+ ix)(d+i x),\qquad
a,b,c,d\in\mathbb{R}_+,\\
\alpha_\mathcal{M}(x) \eqdef
\mathcal{M}\left(\mathcal{M}-1+2(a+b+c+d)\right)x^2.
\end{gather*}
This Hamiltonian can be considered as a deformation by a quadratic
polynomial factor $(a+i x)(b+i x)$ of the exactly solvable
`discrete' quantum mechanics having the continuous Hahn
polynomials as  eigenfunctions~\cite{os5}, another dif\/ference
analogue of the harmonic oscillator. See the comments in Section~5
of~\cite{os10}.

The pseudo ground state wavefunction $\phi_0(x)$ is
\begin{gather*}
\phi_0(x)
\eqdef\sqrt{\Gamma(a+i x)\Gamma(a-i x) \Gamma(b+i x)\Gamma(b-i
x)\Gamma(c+ix)\Gamma(c-i x)\Gamma(d+ix)\Gamma(d-i x)}.
\end{gather*}
Again it has no node and it is square integrable. The similarity
transformed Hamiltonian acting on the polynomial space is
\begin{gather}
  \widetilde{\mathcal{H}} \eqdef
   \phi_0^{-1}\circ\mathcal{H}\circ\phi_0
   =V(x)\big(e^{-i\partial_x}-1\big)+V(x)^*\big(e^{i\partial_x}-1\big)\nonumber\\
   \phantom{\widetilde{\mathcal{H}} \eqdef}{}
 +\mathcal{M}\left(\mathcal{M}-1+2(a+b+c+d)\right)x^2.
   \label{tilH3-2}
\end{gather}
It is straightforward to verify the relationship (\ref{Hactx}) and
to establish the existence of the invariant polynomial subspaces.
In the following subsections, we present Bethe ansatz solutions to
eigenfunctions and eigenvalues of the Hamiltonian~(\ref{tilH3-2}).

\subsubsection[The case of even ${\mathcal M}$]{The case of even $\boldsymbol{{\mathcal M}}$}

Let us introduce a positive integer $N$ such that ${\mathcal
M}=2N$ and a polynomial function $\Psi(x)$ of the form
(\ref{Eign-2-1}). Then $\Psi(x)$ becomes the eigenfunction of
$\widetilde{\mathcal{H}}$ (\ref{tilH3-2}) with an even ${\mathcal
M}$: ${\mathcal M}=2N$ provided that the roots of the polynomial
satisfy the Bethe ansatz equations
\begin{gather}
   \prod_{l\neq j}^{N}
       \frac{(x_j-x_l-i)(x_j+x_l-i)}{(x_j-x_l+i)(x_j+x_l+i)}
 = \frac{V(x_j)^*}{V(x_j)}\,\frac{(\eta(x_j+i)-\eta(x_j))}{(\eta(x_j)-\eta(x_j-i))}
     \label{BAgenform2-2}\\
\qquad{}  = \frac{(a-ix_j)(b-ix_j)(c-ix_j)(d-ix_j)(2x_j+i)}
        {(a+ix_j)(b+ix_j)(c+ix_j)(d+ix_j)(2x_j-i)}, \qquad j= 1,\ldots,N.\nonumber
\end{gather}
The corresponding eigenvalue $E$ is given by
\begin{gather}
   E=2\sum_{j=1}^4{
      {\mathcal M}\choose j}\,\Delta_j-
      \left(4\Delta_3+(4{\mathcal M}-6)\right)
      \sum_{l=1}^Nx_l^2,
\label{Eig-3-2}
\end{gather}
with the binomial coef\/f\/icients
\begin{gather*}
  {\mathcal{M}\choose j}
    =\frac{{\mathcal M}!}{j!\,({\mathcal M}-j)!},
    \qquad j=0,\ldots,{\mathcal M},
\end{gather*}
  and the coef\/f\/icients  $\{\Delta_j|j=1,\ldots,4\}$, which are the elementary
  symmetric polynomials in the parameters $\{a,b,c,d\}$
def\/ined by
\begin{gather}
  V(x)=(a+ix)(b+ix)(c+ix)(d+ix)\eqdef\sum_{j=0}^4\Delta_j(ix)^j,\qquad
\Delta_4=1.
  \label{Par-3-2}
\end{gather}

\subsubsection[The case of odd ${\mathcal M}$]{The case of odd $\boldsymbol{{\mathcal M}}$}

Let us introduce a positive integer $N$ such that ${\mathcal
M}=2N+1$ and a function $\Psi(x)$ of the form~(\ref{Eign-2-2}).
The polynomial $\Psi(x)$ becomes the eigenfunction of
$\widetilde{\mathcal{H}}$ (\ref{tilH3-2}) with an odd ${\mathcal
M}$: ${\mathcal M}=2N+1$ provided that the roots of the polynomial
satisfy the Bethe ansatz equations
\begin{gather}
   \frac{(x_j-i)}{(x_j+i)}\prod_{l\neq j}^{N}
       \frac{(x_j-x_l-i)(x_j+x_l-i)}{(x_j-x_l+i)(x_j+x_l+i)}
 = \frac{V(x_j)^*}{V(x_j)}\,\frac{(\eta(x_j+i)-\eta(x_j))}{(\eta(x_j)-\eta(x_j-i))}
  \label{BAgenform3-2}\\
 \qquad{} = \frac{(a-ix_j)(b-ix_j)(c-ix_j)(d-ix_j)(2x_j+i)}
        {(a+ix_j)(b+ix_j)(c+ix_j)(d+ix_j)(2x_j-i)},\qquad
   j = 1,\ldots,N.\nonumber
\end{gather} The corresponding eigenvalue $E$ is given by the
same expressions as (\ref{Eig-3-2})--(\ref{Par-3-2}) but  with a~dif\/ferent value of ${\mathcal M}$: ${\mathcal M}=2N+1$.

\section{Dif\/ference equation analogues of harmonic oscillator\\ with
centrifugal potential deformed by sextic potential} \label{difsex}

The Bethe ansatz solutions for the dif\/ference equation analogues of the harmonic oscillator with
the centrifugal potential deformed by a sextic potential \cite{deltaqes} are
discussed here. There are two types corresponding to the linear
and quadratic polynomial deformations as discussed in \cite{deltaqes}. The corresponding exactly
solvable dif\/ference equation has the Wilson polynomials
\cite{os4,os5,askey,koeswart} as the eigenfunctions. The
Hamiltonians  have the same form as (\ref{H1}),
(\ref{factformadded}) and (\ref{defAAdag}), with only the
potential function $V(x)$ and the compensation term
$\alpha_\mathcal{M}(x)$ are dif\/ferent:
\begin{gather}
\mbox{Type I}:\quad \ V(x) \eqdef (b+i x)V_0(x),\qquad
\alpha_\mathcal{M}(x)\eqdef
\mathcal{M}x^2, \nonumber\\
\mbox{Type II}:\quad V(x) \eqdef (a+i x)(b+i x)V_0(x),
\label{V-type-II}\\
\phantom{\mbox{Type II}:\quad}{} \ \alpha_\mathcal{M}(x) \eqdef
\mathcal{M}\left(\mathcal{M}-1+(a+b+c+d+e+f)\right)x^2,
\label{A-type-II}
\end{gather}
with a common $V_0(x)$
\begin{gather}
V_0(x) \eqdef \frac{(c+ ix)(d+i x)(e+i x)(f+i x)}{2ix(2ix +1)},
\qquad a,b,c,d,e,f\in\mathbb{R}_+-\{{1}/{2}\}. \label{V0Wilson}
\end{gather}
None of the parameters $a$, $b$, $c$, $d$, $e$ or $f$ should take
the value $1/2$, since it would cancel the denominator. Because of
the centrifugal barrier, the dynamics is constrained to a half
line; $0<x<\infty$. The type I case can also be considered as a
quadratic polynomial deformation of  the exactly solvable dynamics
with  $V_{01}(x)$:
\begin{gather}
\mbox{Type I}:\quad V(x) \eqdef (b+i x)(c+i x)V_{01}(x),\qquad
V_{01}(x)\eqdef\frac{(d+i x)(e+i x)(f+i x)}{2ix(2ix +1)},
\label{V0condualHahn}
\end{gather}
which has the continuous dual Hahn polynomials
\cite{os4,os5,askey,koeswart} as eigenfunctions. This
re-inter\-pre\-ta\-tion does not change the dynamics, since the
Hamiltonian and~$A$ and~$A^\dagger$ operators depend on~$V(x)$.

\subsection{Type I theory}
Here we consider the dif\/ference equation of type I. The  pseudo
ground state wavefunction $\phi_0(x)$ is determined as the zero
mode of  the $A$ operator \eqref{defAAdag}, $A\phi_0=0$:
\begin{gather*}
  \mbox{Type I}:\quad \phi_0(x) \eqdef  \frac{
  \sqrt{\prod\limits_{j=2}^6 \Gamma(a_j+i x)\Gamma(a_j-i x)}}{\sqrt{\Gamma(2i
   x)\Gamma(-2ix)}},
\end{gather*}
in which the numbering of the parameters
\begin{gather}
a_1\eqdef a,\qquad a_2\eqdef b,\qquad a_3\eqdef c,\qquad a_4\eqdef d,\qquad a_5\eqdef
e,\qquad a_6\eqdef f, \label{label-para}
\end{gather}
is used. It is obvious that  $\phi_0$ has no node in the half line
$0<x<\infty$.

The similarity transformed Hamiltonian acting on the polynomial
space has the same form as before (\ref{tilH3})
\begin{gather}
  \widetilde{\mathcal{H}} \eqdef
   \phi_0^{-1}\circ\mathcal{H}\circ\phi_0
   =V(x)\big(e^{-i\partial_x}-1\big)+V(x)^*\big(e^{i\partial_x}-1\big)
+\mathcal{M} \eta(x).
   \label{tilH4}
\end{gather}
Again the Hamiltonian is  parity invariant, that is
\begin{gather*}
\left.\mathcal{H}\right|_{x\to -x}=\mathcal{H}.
\end{gather*}
Although the potential $V(x)$ has the harmful looking denominator
$1/\{2ix(2ix+1)\}$, it is straightforward to verify that
$\widetilde{\mathcal{H}}$ maps a polynomial in $\eta(x)=x^2$ into
another, as $2ix+1\propto \eta(x-i)-\eta(x)$:
\begin{gather*}
\widetilde{\mathcal{H}} x^{2n}=\left\{\begin{array}{lll}
\displaystyle \sum_{j=0}^{n+1}a_{n, j}x^{2n+2-2j},& n\leq \mathcal{M}-1,&
a_{n,\,j}\in\mathbb{R},\vspace{1mm}\\
\displaystyle \sum_{j=0}^{\mathcal{M}}a'_{n, j}x^{2\mathcal{M}-2j},&  n=
\mathcal{M},&a'_{n, j}\in\mathbb{R}.
\end{array}\right.
\end{gather*}
This is because $V_0$, which has the above denominator, keeps the
polynomial subspace of any even degree invariant, ref\/lecting the
exact solvability. In other words, the exactly solvable discrete
quantum mechanics corresponding to the undeformed potentials
$V_0(x)$ or $V_{01}(x)$ has the eigenfunction
\[
\phi_0(x)P(\eta(x)),\qquad \eta(x)=x^2,
\]
in which $P(\eta)$ is either the continuous dual Hahn polynomial
($V_{01}(x)$, \eqref{V0condualHahn}), or the Wilson polynomial
($V_0(x)$, \eqref{V0Wilson}). This establishes that
$\widetilde{\mathcal{H}}$ keeps the polynomial space $ {\mathcal
V}_{\mathcal M}$ invariant,
\begin{gather}
  \widetilde{\mathcal{H}} {\mathcal V}_{\mathcal M} \subseteq
   {\mathcal V}_{\mathcal M},\nonumber\\
     {\mathcal V}_{\mathcal M}
   \eqdef
    \mbox{Span}\big[1,x^2,\ldots,x^{2k},\ldots,x^{2\mathcal
     M}\big],\qquad
    \mbox{dim}{\mathcal V}_{\mathcal M}= {\mathcal M}+1.
    \label{subspc3-1}
\end{gather}
The above equations imply that the eigenfunctions of
$\widetilde{\mathcal{H}}$ in the subspace ${\mathcal V}_{\mathcal
M}$ are of the form
\begin{gather*}
  \Psi(x)=\prod_{l=1}^{\mathcal
M}\left(\eta(x)-\eta(x_l)\right)=\prod_{l=1}^{\mathcal
M}(x-x_l)(x+x_l).
\end{gather*}
Analogous calculation shows that the polynomial $\Psi(x)$ becomes
the eigenfunction of $\widetilde{\mathcal{H}}$ if the roots of the
polynomial  satisfy the Bethe ansatz equations
\begin{gather}
  \prod_{l\neq j}^{{\mathcal M}}
    \frac{(x_j-x_l-i)(x_j+x_l-i)}{(x_j-x_l+i)(x_j+x_l+i)}
 =\frac{V(x_j)^*}{V(x_j)}\,\frac{\eta(x_j+i)-\eta(x_j)}{\eta(x_j)-\eta(x_j-i)}
   \label{BAgenform4}\\
\qquad{} =\frac{(b-ix_j)(c-ix_j)(d-ix_j)(e-ix_j)(f-ix_j)}
      {(b+ix_j)(c+ix_j)(d+ix_j)(e+ix_j)(f+ix_j)},\qquad
  j =1,\ldots, {\mathcal M}.
  \label{BAE-3}
\end{gather}
Note that the kinematical factors $\pm 2ix+1$ of $V(x)$ and
$V(x)^*$ are cancelled by $\eta(x\mp i)-\eta(x)$. The
corresponding eigenvalue $E$ is given by
\begin{gather}
   E = \frac{2}{3}{\mathcal M}({\mathcal M}-1)({\mathcal M}-2)
      +\left(b+c+d+e+f+\frac{1}{2}\right){\mathcal M}({\mathcal M}-1)\nonumber\\
\phantom{E=}+\left\{b(c+d+e+f)+c(d+e+f)+d(e+f)+ef\right\}{\mathcal M}
     -\sum_{l=1}^{{\mathcal M}}x^2_l,\label{Eigenvalue-3}
\end{gather}
where $\{x_l\}$ satisfy the Bethe ansatz equations~(\ref{BAE-3}).

As expected, exactly solvable limits are obtained by two dif\/ferent
ways; either one or two parameters go to inf\/inity. In the former
case, the scaled Hamiltonian \eqref{H1} $\mathcal{H}/f$ gives that
of the Wilson polynomials with four real parameters, $b$, $c$, $d$
and $e$. The eigenvalue formula \eqref{Eigenvalue-3} reduces to
that of the Wilson polynomials \cite{os4,os5,os13}
\begin{equation}
\lim_{f\to\infty}E/f =\mathcal{M}(\mathcal{M}+b+c+d+e-1).
\label{eformWil}
\end{equation}
In the latter case ($e,f\to\infty$), the scaled Hamiltonian
\eqref{H1} $\mathcal{H}/(ef)$ gives that of the continuous dual
Hahn polynomials with  three parameters, $b$, $c$ and $d$. The
eigenvalue formula \eqref{Eigenvalue-3} reduces to that of the
continuous dual Hahn polynomials \cite{os4,os5,os13}
\begin{equation}
\lim_{e,\,f\to\infty}E/(ef)=\mathcal{M}. \label{eformcdH}
\end{equation}
The Bethe ansatz equations \eqref{BAE-3} in these limits determine
the zeros of the Wilson and the continuous dual Hahn polynomials,
respectively:
\begin{gather}
 f\to\infty:\quad \ \  \;  \prod_{l\neq j}^{{\mathcal M}}
    \frac{(x_j-x_l-i)(x_j+x_l-i)}{(x_j-x_l+i)(x_j+x_l+i)}
       =\frac{(b-ix_j)(c-ix_j)(d-ix_j)(e-ix_j)}
      {(b+ix_j)(c+ix_j)(d+ix_j)(e+ix_j)},\label{rBAE-W}
      \\
  e, f\to\infty:\quad \prod_{l\neq j}^{{\mathcal M}}
    \frac{(x_j-x_l-i)(x_j+x_l-i)}{(x_j-x_l+i)(x_j+x_l+i)}
       =\frac{(b-ix_j)(c-ix_j)(d-ix_j)}
      {(b+ix_j)(c+ix_j)(d+ix_j)}. \label{rBAE-3}
 \end{gather}

\subsection{Type II theory}
Here we consider the dif\/ference equation of type II. The pseudo
ground state wavefunction $\phi_0(x)$ is determined again as the
zero mode of  the $A$ operator~\eqref{defAAdag}, $A\phi_0=0$:
\begin{gather*}
\mbox{Type II}:\quad \phi_0(x) \eqdef
    \frac{\sqrt{\prod_{j=1}^6\Gamma(a_j+i x)
    \Gamma(a_j-ix)}}{\sqrt{\Gamma(2i x)\Gamma(-2ix)}},
\end{gather*}
where the same numbering of the parameters as those in
(\ref{label-para}) has been used. The similarity transformed
Hamiltonian acting on the polynomial space has the similar form as
(\ref{tilH4}) but with a dif\/ferent potential function $V(x)$
(\ref{V-type-II}) and compensation term $\alpha_{{\mathcal M}}(x)$
(\ref{A-type-II})
\begin{gather}
   \widetilde{\mathcal{H}} \eqdef
    \phi_0^{-1}\circ\mathcal{H}\circ\phi_0
    =V(x)\big(e^{-i\partial_x}-1\big)+V(x)^*\big(e^{i\partial_x}-1\big)
    \nonumber\\
\phantom{\widetilde{\mathcal{H}} \eqdef
    \phi_0^{-1}\circ\mathcal{H}\circ\phi_0=}{}
    +\mathcal{M}\left(\mathcal{M}-1+(a+b+c+d+e+f)\right)x^2.
    \label{tilH4-2}
\end{gather} It can be verif\/ied   that the resulting Hamiltonian
$\widetilde{\mathcal{H}}$ (\ref{tilH4-2})  keeps the polynomial
space $ {\mathcal V}_{\mathcal M}$ def\/ined in (\ref{subspc3-1})
invariant. Like the type~I case, this allows us to search the
corresponding eigenfunction~$\Psi(x)$ of form
$\Psi(x)=\prod_{l=1}^{\mathcal M}(x-x_l)(x+x_l)$. Analogous
calculation shows that the polynomial~$\Psi(x)$ becomes the
eigenfunction of~$\widetilde{\mathcal{H}}$ (\ref{tilH4-2})
provided that the roots of the polynomial  satisfy the Bethe
ansatz equations
\begin{gather}
   \prod_{l\neq j}^{{\mathcal M}}
     \frac{(x_j-x_l-i)(x_j+x_l-i)}{(x_j-x_l+i)(x_j+x_l+i)}
 = \frac{V(x_j)^*}{V(x_j)}\,\frac{(\eta(x_j+i)-\eta(x_j))}{(\eta(x_j)-\eta(x_j-i))}
         \label{BAgenform4-2}\\
\qquad{}= \frac{(a-ix_j)(b-ix_j)(c-ix_j)(d-ix_j)(e-ix_j)(f-ix_j)}
       {(a+ix_j)(b+ix_j)(c+ix_j)(d+ix_j)(e+ix_j)(f+ix_j)},\qquad
 j=1,\ldots, {\mathcal M}.
       \label{BAE-3-2}
\end{gather}
The corresponding eigenvalue $E$ is given by
\begin{gather*}
E =\Delta_3{{\mathcal M}\choose1}
    +(2\Delta_4+\Delta_5){{\mathcal M}\choose 2}
    +4(\Delta_5+1){{\mathcal M}\choose3}
    +8{{\mathcal
    M}\choose4}
 -\left(\Delta_5+2({\mathcal M}-1)\right)
   \! \sum_{l=1}^{{\mathcal M}}x_l^2.\!
\end{gather*}
Here the coef\/f\/icients $\{\Delta_j|j=1,\ldots,6\}$ are the
elementary symmetric polynomials in the para\-me\-ters
$\{a,b,c,d,e,f\}$ def\/ined by
\begin{gather*}
(a+ix)(b+ix)(c+ix)(d+ix)(e+ix)(f+ix)\eqdef\sum_{j=0}^6\Delta_j(ix)^j,
\qquad \Delta_6=1.
\end{gather*}
Various limits to the exactly solvable cases are almost the same
as in the type I theory and will not be listed here.

It is interesting to note that the type II Hamiltonian of Section~\ref{difex1},  \eqref{H1}, \eqref{typeIIV}, \eqref{typeIIVa},  is
obtained from the type II Hamiltonian of Section~\ref{difsex},
   \eqref{H1},
\eqref{V-type-II}, \eqref{A-type-II} as a formal limit $e\to0$,
$f\to1/2$:
\begin{gather*}
4\mathcal{H}_{\text{Section~4}}|_{e\to0,f\to1/2}=\mathcal{H}_{\text{Section~3}}|_{\mathcal{M}\to2\mathcal{M}}.
\end{gather*}
The Bethe ansatz equations together with the eigenvalue formulas
are related in similar ways.

\section[Difference equation analogue of $1/\sin^2x$ potential deformed by
$\cos2x$ potential]{Dif\/ference equation analogue of $\boldsymbol{1/\sin^2x}$ potential\\ deformed by
$\boldsymbol{\cos2x}$ potential} \label{diftrig}

  The last example is the dif\/ference analogue of the model discussed in
Subsection~2.1.2 of \cite{deltaqes}, $1/\sin^2x$ potential
deformed by a $\cos2x$ potential. In this case the corresponding
exactly solvable dif\/ference equation has the Askey--Wilson
polynomials \cite{os4,os5,askey,koeswart} as  eigenfunctions. The
basic idea for showing quasi exact solvability is almost the same
as shown above.

As introduced and explored in \cite{deltaqes}, this system is a
quasi exactly solvable deformation of the exactly solvable
dynamics which has the Askey--Wilson polynomials
\cite{os4,os5,askey,koeswart} as  eigenfunctions. Here we slightly
change the notation from that of \cite{deltaqes} for consistency
with the rest of this paper. The range of the parameter $x$ is now
f\/inite, to be chosen as
\begin{gather*}
   0<x<\pi,
   \end{gather*}
and we introduce a complex variable $z$ and the sinusoidal
coordinate $\eta(x)$:
\begin{gather*}
z=e^{i x},\qquad \eta(x)\eqdef \cos x=(z+z^{-1})/2.
\end{gather*}
The unit of the shift is changed from 1 to a real constant
$\gamma\eqdef\log q$, $0<q<1$. Then the shift operator $e^{\gamma
p}$ can be written as
\begin{gather*}
   e^{\gamma p}=e^{-i\gamma\frac{d}{dx}}=q^D,\qquad D\eqdef{z\frac{d}{dz}},
\end{gather*}
whose action on a function of $x$ can be expressed as $z\to qz$:{\samepage
\[
   e^{\gamma
p}f(x)=f(x-i\gamma)=q^D\check{f}(z)=q^{z\frac{d}{dz}}\check{f}(z)=\check{f}(qz),
   \qquad \text{with}\quad  f(x)=\check{f}(z).
\]
Note that $\gamma<0$.}

The  Hamiltonian takes the form
\begin{gather}
   \mathcal{H} \eqdef\sqrt{V(x)}\,e^{\gamma p}\sqrt{V(x)^*}
   +\sqrt{V(x)^*}\,e^{-\gamma p}\sqrt{V(x)}-V(x)-V(x)^*+\alpha_\mathcal{M}(x),
  \nonumber \\
  \phantom{\mathcal{H}}{} \, =\sqrt{V(x)}\,q^{D}\!\sqrt{V(x)^*}
   +\sqrt{V(x)^*}\,q^{-D}\!\sqrt{V(x)}
   -(V(x)+V(x)^*)+\alpha_\mathcal{M}(x),
   \label{H-q}\\
 \phantom{\mathcal{H}}{} \, =A^\dagger A+\alpha_\mathcal{M}(x),\qquad
\alpha_\mathcal{M}(x)\eqdef -2abcde
q^{-1}\big(1-q^\mathcal{M}\big)\eta(x),\nonumber\\
{A}^{\dagger} \eqdef
   -i\left(\sqrt{V(x)}\,q^{\frac{D}{2}}
   -\sqrt{V(x)^*}\,q^{-\frac{D}{2}}\right),\qquad
{A}\eqdef
   i\left(q^{\frac{D}{2}}\sqrt{V(x)^*}
   -q^{-\frac{D}{2}}\sqrt{V(x)}\right),
   \label{qAdef}\\
V(x) \eqdef(1-a z)V_0(x),\qquad
V_0(x)\eqdef\frac{(1-bz)(1-cz)(1-dz)(1-ez)}{(1-z^2)(1-qz^2)},\nonumber\\
 -1<a,b,c,d,e<1. \label{pararange}
\end{gather}
The Hamiltonian is obtained by deforming the potential function
$V_0(z)$ by a linear polynomial in~$z$. The parameter range
\eqref{pararange} could be enlarged to one real parameter (say,
$a$) and two complex conjugate pairs (for example, $b=c^*$,
$d=e^*$), but the absolute values must be less than~1, $|a|<1$,
\ldots, $|e|<1$.

The  pseudo ground state wavefunction $\phi_0(x)$ is determined as
the zero mode of  the $A$ opera\-tor~\eqref{qAdef}, $A\phi_0=0$:
\begin{gather*}
  \phi_0(x)\eqdef
\sqrt{\frac{(z^2,z^{-2};q)_{\infty}}
   {(az,az^{-1},bz,bz^{-1},cz,cz^{-1},dz,dz^{-1},ez,ez^{-1};q)_{\infty}}},
\end{gather*}
where $(a_1,\dots,a_m;q)_{\infty}\eqdef\prod_{j=1}^m
\prod_{n=0}^{\infty}(1-a_jq^n)$. Obviously $\phi_0$ has no node or
singularity in $0<x<\pi$. We look for exact eigenvalues and
eigenfunctions of the  Hamiltonian~(\ref{H-q}) in the form:
\begin{gather*}
{\mathcal H}\phi=E\phi,\qquad \phi(x)=\phi_0(x)\Psi(\eta(x)),\qquad
\eta(x)=\cos x=(z+z^{-1})/2, 
\end{gather*}
in which $\Psi(\eta(x))$ is a  polynomial in $\eta(x)$ or in
$(z+{1}/{z})/2=\cos x$. The similarity transformed Hamiltonian
acting on the polynomial space has the form
\begin{gather*}
   \widetilde{\mathcal{H}}\eqdef
   \phi_0^{-1}\circ \mathcal{H}\circ\phi_0
   =V(z)\left(q^{D}-1\right)+V(z)^*\left(q^{-D}-1\right)
   -abcde q^{-1}\big(1-q^\mathcal{M}\big)\left(z+\frac{1}{z}\right).
\end{gather*}
Without the deformation factor $1-az$ and the compensation term,
the above Hamiltonian $\widetilde{\mathcal{H}}$ is exactly
solvable, that is, it keeps the polynomial subspace in
$\eta(x)=(z+{1}/{z)/2}$ of any degree invariant. The deformed
Hamiltonian $\widetilde{\mathcal{H}}$ is parity invariant
$\left.\mathcal{H}\right|_{x\to -x}=\mathcal{H}$ and it is
straightforward to show the existence of an invariant polynomial
subspace:
\begin{gather*}
\widetilde{\mathcal{H}} {\mathcal V}_{\mathcal M} \subseteq
{\mathcal V}_{\mathcal M},
\\
     {\mathcal V}_{\mathcal M}
 \eqdef  \mbox{Span}\big[1,\eta(x),\ldots,\eta(x)^k,\ldots,
\eta(x)^\mathcal{M}\big],\qquad \dim {\mathcal V}_{\mathcal M}=
{\mathcal M}+1. 
\end{gather*}
The above equations imply that the eigenfunctions of
$\widetilde{\mathcal{H}}$ in the subspace ${\mathcal V}_{\mathcal
M}$ are of the form
\begin{gather*}
  \Psi(x)=\prod_{l=1}^{{\mathcal M}}(\eta(x)-\eta(x_l))
  =\prod_{l=1}^{{\mathcal M}}(\cos x-\cos x_l)
  \equiv P_{\mathcal M}(\cos x).
\end{gather*}
Analogous calculation shows that $\Psi(x)$ becomes the
eigenfunction of $\widetilde{\mathcal{H}}$ if the parameters
$\{x_l\}$ satisfy the Bethe ansatz equations
\begin{gather}
\prod_{l\neq j}^{{\mathcal M}}
    \frac{\cos(x_j-i\gamma)-\cos x_l}
     {\cos(x_j+i\gamma)-\cos x_l} =
\frac{V(x_j)^*}{V(x_j)}\,\frac{\eta(x_j+i\gamma)-\eta(x_j)}{\eta(x_j)-\eta(x_j-i\gamma)}  \label{BAgenform5}\\
\qquad{} =\frac{(z_j-a)(z_j-b)(z_j-c)(z_j-d)(z_j-e)}
     {(1-az_j)(1-bz_j)(1-cz_j)(1-dz_j)(1-ez_j)z_j},\qquad j=1,\ldots, {\mathcal M}.
\label{BAE-4}
\end{gather}
Note that  $z_j=e^{i x_j}$ and
$\eta(x_j-i\gamma)=\cos(x_j-i\gamma)=(qz+q^{-1}z^{-1})/2$, etc.
Again the kinematical factors $(1-z^{\pm2})(1-qz^{\pm2})$ of
$V(x)$ and $V(x)^*$ are cancelled by $\eta(x\mp i\gamma)-\eta(x)$.
The corresponding eigenvalue $E$ is given by
\begin{gather}
   E = (abcd+abce+abde+acde+bcde) q^{-1}\big(q^{{\mathcal M}}-1\big)
    +q^{-{\mathcal M}}-1\nonumber\\
\phantom{E=}-2abcde\,q^{{\mathcal M}-1}\big(1-q^{-1}\big)\sum_{l=1}^{{\mathcal M}}
     \cos x_l,\label{Eigenvalue-4}
\end{gather}
where $\{x_l\}$ satisfy the Bethe ansatz equations (\ref{BAE-4}).

In this example, there are many ways to obtain exactly solvable
dynamics; by making either one ($e$), two ($d,e$), three
($c,d,e$), four ($b,c,d,e$) or f\/ive ($a,b,c,d,e$) parameters
vanish. The corresponding Hamiltonians are those describing the
dynamics of the Askey--Wilson, continuous dual $q$-Hahn,
Al-Salam--Chihara, big $q$-Hermite and $q$-Hermite polynomials,
respectively \cite{askey,koeswart,os4,os5,os13}. The eigenvalue
formula \eqref{Eigenvalue-4} reduces to that of the Askey--Wilson
polynomials for $e=0$ \cite{os4,os5,os13};
\begin{gather}
E=\big(q^{-\mathcal M}-1\big)\big(1-abcd q^{\mathcal M+1}\big), \label{eformAW}
\end{gather}
and to a universal formula
\begin{gather}
E=q^{-\mathcal M}-1, \label{eformrest}
\end{gather}
for the rest  \cite{os4,os5,os13}. The Bethe ansatz equations
\eqref{BAE-4} for the restricted case of $e=0$
\begin{gather}
\prod_{l\neq j}^{{\mathcal M}}
    \frac{\cos(x_j-i\gamma)-\cos x_l}
     {\cos(x_j+i\gamma)-\cos x_l} =\frac{(z_j-a)(z_j-b)(z_j-c)(z_j-d)}
     {(1-az_j)(1-bz_j)(1-cz_j)(1-dz_j)},
       \qquad j=1,\ldots, {\mathcal M},
\label{rBAE-AW}
\end{gather}
then determine the zeros of the Askey--Wilson polynomials. Further
restrictions $d,e=0$, $c,d,e=0$, $b,c,d,e=0$ and $a,b,c,d,e=0$
determine the zeros of the continuous dual $q$-Hahn,
Al-Salam--Chihara, big $q$-Hermite and $q$-Hermite polynomials,
respectively~\cite{koeswart,os13}.

\section{Summary and comments}

We have constructed Bethe ansatz solutions for the quasi exactly
solvable dif\/ference equations of one degree of freedom
introduced in \cite{deltaqes} and
\cite{newqes}.  These quasi exactly solvable dif\/ference equations are
deformations of the well-known exactly solvable dif\/ference
equations of the Meixner--Pollaczek, continuous Hahn, continuous
dual Hahn, Wilson and Askey--Wilson polynomials. The eigenfunctions
within the exactly solvable subspace are explicitly given by some
polyno\-mials (module a pseudo ground state wavefunction $\phi_0$)
whose roots are solutions of the associated Bethe ansatz
equations. The corresponding eigenvalues are expressed in terms of
the solutions of the Bethe ansatz equations. These are dif\/ference
equation counterparts of the results of Sasaki--Takasaki
\cite{st1}, which gave a Bethe ansatz formulation of the QES
systems corresponding to the harmonic oscillator (with/without a
centrifugal barrier) deformed by a sextic potential and the
$1/\sin^2x$ potential deformed by a $\cos2x$ potential.

As in the exactly solvable quantum mechanics, the sinusoidal
coordinates $\eta(x)$ \cite{os7,os9} play an essential role. The
exactly solvable sector is spanned by it
\begin{gather*}
\mbox{Span}\big[1,\eta(x),\ldots,\eta(x)^{k},\ldots,\eta(x)^{\mathcal
     M}\big],
\end{gather*}
then the Bethe ansatz equations take an almost universal form,
\eqref{BAgenform1}, \eqref{BAgenform2}, \eqref{BAgenform3},
\eqref{BAgenform2-2}, \eqref{BAgenform3-2}, \eqref{BAgenform4},
\eqref{BAgenform4-2} and \eqref{BAgenform5}. The limits or
restrictions to the exactly solvable dynamics are demonstrated in
detail including various eigenvalue formulas \eqref{eformCH},
\eqref{eformMP}, \eqref{eformcH2}, \eqref{eformMP2},
\eqref{eformWil}, \eqref{eformcdH}, \eqref{eformAW} and
\eqref{eformrest}. The Bethe ansatz equations reduce to those
determining the zeros of the corresponding orthogonal polynomials,
for example \eqref{rBAE-MP}, \eqref{rBAE-W}, \eqref{rBAE-3} and
\eqref{rBAE-AW}.

All the known quasi exactly solvable dynamics have the exactly
solvable subspace consisting of polynomials in a certain variable
(see, e.g.~\cite{turb2}). Our emphasis here is that the variable is the
sinusoidal coordinate which plays the central role in the
corresponding exactly solvable limits~\cite{os7, os9, os13}.

It should be mentioned that there exist some examples of deriving (quasi) exactly solvable
dif\/ference equations in terms of Lie algebraic deformations of exactly solvable dynamics
introduced by Turbiner and his collaborators \cite{turb3}.
These dif\/ference equations have shifts in the real direction $\psi(x\pm1)$ and the
corresponding eigenfunctions have discrete orthogonality measures, in contrast
to those discussed in this paper which have pure imaginary shifts, $\psi(x\pm i)$ or
$\psi(x\pm i\gamma)$, $\gamma\in{\mathbb R}$, and the corresponding eigenfunctions have
continuous orthogonality measures.

The deformations of exactly solvable dynamics for obtaining the
quasi exactly solvable quantum systems introduced in
\cite{st1,deltaqes,newqes} and discussed in this paper in detail,
are of the simplest type, in which the compensation term is linear
in the sinusoidal coordinate. Possibility of further deformations
including quadratic compensation terms will be discussed in
\cite{os14}, in particular, for those quantum systems having
discrete orthogonality measures \cite{os12}.


\section*{Acknowledgements}

The f\/inancial support from  Australian Research Council  is
gratefully acknowledged. R.S.\ is supported in part by
Grants-in-Aid for Scientif\/ic Research from the Ministry of
Education, Culture, Sports, Science and Technology, No.18340061
and No.19540179. Y.Z.Z.\ thanks the Yukawa Institute for
Theoretical Physics, Kyoto University for hospitality and
f\/inancial support.

\pdfbookmark[1]{References}{ref}
\LastPageEnding


\begin{thebibliography}{99}

\footnotesize\itemsep=0pt


\bibitem{susyqm}
Infeld L., Hull  T.E.,
The factorization method,
{\it Rev. Modern Phys.} {\bf 23} (1951), 21--68.\\
Cooper F., Khare A., Sukhatme U.,
Supersymmetry and quantum mechanics,
{\it Phys. Rep.} {\bf 251} (1995), 267--385,
\href{http://arxiv.org/abs/hep-th/9405029}{hep-th/9405029}.

\bibitem{os4}
Odake S., Sasaki R.,
Shape invariant potentials in ``discrete quantum mechanics'',
{\it J. Nonlinear Math. Phys.} {\bf 12} (2005), suppl.~1, 507--521,
\href{http://arxiv.org/abs/hep-th/0410102}{hep-th/0410102}.

\bibitem{os5}
Odake  S., Sasaki  R.,
Equilibrium positions, shape invariance and Askey--Wilson polynomials,
{\it J. Math. Phys.} {\bf 46} (2005),  063513, 10~pages,
\href{http://arxiv.org/abs/hep-th/0410109}{hep-th/0410109}.\\
Odake  S., Sasaki  R., Calogero--Sutherland--Moser systems, Ruijsenaars--Schneider--van~Diejen systems and orthogonal polynomials,
{\it Prog. Theoret. Phys.} {\bf 114} (2005), 1245--1260,
\href{http://arxiv.org/abs/hep-th/0512155}{hep-th/0512155}.\\
Odake  S., Sasaki  R., Equilibrium positions and eigenfunctions of shape invariant
(``discrete'') quantum mechanics,
{\it Rokko Lectures in Mathematics (Kobe University)} {\bf 18} (2005), 85--110,
\href{http://arxiv.org/abs/hep-th/0505070}{hep-th/0505070}.

\bibitem{os7}
Odake  S., Sasaki  R.,
Unif\/ied theory of annihilation-creation operators for solvable (``discrete'') quantum mechanics,
{\it J. Math. Phys.} {\bf 47} (2006), 102102, 33~pages,
\href{http://arxiv.org/abs/quant-ph/0605215}{quant-ph/0605215}.\\
Odake  S., Sasaki  R.,
Exact solution in the Heisenberg picture and annihilation-creation operators,
{\it Phys. Lett. B} {\bf 641} (2006), 112--117,
\href{http://arxiv.org/abs/quant-ph/0605221}{quant-ph/0605221}.

\bibitem{os9}
Odake  S., Sasaki  R.,
Exact Heisenberg operator solutions for multi-particle quantum mechanics,
{\it J. Math. Phys.} {\bf 48} (2007), 082106, 12~pages,
\href{http://arxiv.org/abs/0706.0768}{arXiv:0706.0768}.

\bibitem{os13}
Odake  S., Sasaki  R.,
Exactly solvable `discrete' quantum mechanics; shape invariance, Heisenberg solutions, annihilation-creation operators and coherent states,
{\it Prog. Theoret. Phys.} {\bf 119} (2008), 663--700,
\href{http://arxiv.org/abs/0802.1075}{arXiv:0802.1075}.


\bibitem{os12}
Odake  S., Sasaki  R.,
Orthogonal polynomials from Hermitian matrices,
{\it J. Math. Phys.} {\bf 49} (2008),  053503, 43~pages,
\href{http://arxiv.org/abs/0712.4106}{arXiv:0712.4106}.


\bibitem{askey}
Andrews G.E., Askey R., Roy  R.,
Special functions, {\it Encyclopedia of Mathematics and Its Applications}, Vol.~71,
Cambridge University Press, Cambridge, 1999.

\bibitem{koeswart}
Koekoek  R., Swarttouw  R.F.,
The Askey-scheme of hypergeometric orthogonal polynomials and its $q$-analogue,
\href{http://arxiv.org/abs/math.CA/9602214}{math.CA/9602214}.

\bibitem{deltaqes}
Sasaki  R.,
Quasi exactly solvable dif\/ference equations,
{\it J. Math. Phys.} {\bf 48} (2007), 122104, 11 pages,
\href{http://arxiv.org/abs/0708.0702}{arXiv:0708.0702}.

\bibitem{newqes}
Sasaki  R.,
New quasi exactly solvable dif\/ference equation,
{\it J. Nonlinear Math. Phys.}
{\bf 15}  (2008), suppl.~3,  373--384,
\href{http://arxiv.org/abs/0712.2616}{arXiv:0712.2616}.




\bibitem{Ush}
Ushveridze  A.G., Quasi-exactly solvable models in quantum mechanics,
Institute of Physics Publishing, Bristol, 1994.\\
Morozov A.Y., Perelomov  A.M., Roslyi  A.A., Shifman  M.A., Turbiner  A.V.,
Quasi-exactly-solvable quantal problems: one-dimensional analog of rational conformal f\/ield theories,
{\it Internat. J. Modern Phys. A} {\bf 5} (1990), 803--832.


\bibitem{turb}
Turbiner  A.V.,
Quasi-exactly-solvable problems and ${\rm sl}(2)$ algebra,
{\it Comm. Math. Phys.} {\bf 118} (1988), 467--474.


\bibitem{N-SUSY}
Andrianov  A.A.,  Iof\/fe  M.V., Spiridonov  V.P.,
Higher-derivative supersymmetry and the Witten index,
{\it Phys. Lett.~A} {\bf 174} (1993), 273--279,
\href{http://arxiv.org/abs/hep-th/9303005}{hep-th/9303005}.\\
 Bagrov  V.G., Samsonov  B.F.,
 Darboux transformation, factorization and supersymmetry in one-dimensional quantum mechanics,
 {\it Theoret. and Math. Phys.} {\bf 104} (1995), 1051--1060.\\
  Klishevich  S.M., Plyushchay  M.S.,
  Supersymmetry of parafermions,
  {\it Modern Phys. Lett. A} {\bf 14} (1999), 2739--2752,
\href{http://arxiv.org/abs/hep-th/9905149}{hep-th/9905149}.\\
   Aoyama  H., Kikuchi  H., Okouchi  I., Sato  M., Wada  S.,
   Valley views: instantons, large order  behaviors, and supersymmetry,
   {\it Nuclear Phys. B} {\bf 553}  (1999), 644--710,
\href{http://arxiv.org/abs/hep-th/9808034}{hep-th/9808034}.\\
  Aoyama  H., Sato  M., Tanaka  T.,
  General forms of a ${\mathcal N}$-fold supersymmetric family,
  {\it Phys. Lett. B} {\bf 503} (2001), 423--429,
\href{http://arxiv.org/abs/quant-ph/0012065}{quant-ph/0012065}.

\bibitem{st1}
Sasaki  R., Takasaki  K.,
Quantum Inozemtsev model, quasi-exact solvability and ${\cal N}$-fold supersymmetry,
{\it J.~Phys.~A: Math. Gen.} {\bf 34} (2001), 9533--9553,
Corrigendum, {\it J. Phys. A: Math. Gen.} {\bf 34} (2001), 10335,
\href{http://arxiv.org/abs/hep-th/0109008}{hep-th/0109008}.


\bibitem{os10}
Odake  S., Sasaki  R.,
Multi-particle quasi exactly solvable dif\/ference equations,
{\it J. Math. Phys.} {\bf 48} (2007), 122105,  8~pages,
\href{http://arxiv.org/abs/0708.0716}{arXiv:0708.0716}.


\bibitem{Wiegmann}
Wiegmann    P.B., Zabrodin  A.V.,
Bethe-ansatz for Bloch electron in magnetic f\/ield,
{\it Phys. Rev. Lett.} {\bf 72} (1994), 1890--1893.\\
Wiegmann    P.B., Zabrodin  A.V.,
Algebraization of dif\/ference eigenvalue equations related to $U_q({\rm sl}_2)$,
{\it Nuclear Phys. B} {\bf 451} (1995), 699--724,
\href{http://arxiv.org/abs/cond-mat/9501129}{cond-mat/9501129}.

\bibitem{Fel96}
Felder  G., Varchenko A.,
Algebraic Bethe ansatz for the elliptic quantum group $E_{\tau,\eta}(sl_2)$,
{\it Nuclear Phys. B} {\bf 480} (1996), 485--503,
\href{http://arxiv.org/abs/q-alg/9605024}{q-alg/9605024}.

\bibitem{Hou03}
Hou  B.Y., Sasaki R., Yang W.-L.,
Algebraic Bethe ansatz for the elliptic quantum group $E_{\tau,\eta}({\rm sl}_n)$ and its applications,
{\it Nuclear Phys. B} {\bf 663} (2003), 467--486,
\href{http://arxiv.org/abs/hep-th/0303077}{hep-th/0303077}.\\
Hou  B.Y., Sasaki R., Yang W.-L.,
Eigenvalues of Ruijsenaars--Schneider model associated with $A_{n-1}$ root system in Bethe ansatz formalism,
{\it J.\ Math.\ Phys.} {\bf 45} (2004), 559--575,
\href{http://arxiv.org/abs/hep-th/0309194}{hep-th/0309194}.

\bibitem{Man07}
Manojlovic  N., Nagy  Z.,
Construction of the Bethe state for the $E_{\tau,\eta}({\rm so}(3))$ elliptic quantum group,
{\it SIGMA} {\bf 3}  (2007), 004, 10~pages,
\href{http://arxiv.org/abs/math.QA/0612086}{math.QA/0612086}.\\
Manojlovic  N., Nagy  Z.,
Algebraic Bethe ansatz for the elliptic quantum group $E_{\tau,\eta}(A_2^{(2)})$,
{\it J. Math. Phys.} {\bf 48} (2007), 123515, 11~pages,
\href{http://arxiv.org/abs/0704.3032}{arXiv:0704.3032}.


\bibitem{degruij}
Degasperis  A., Ruijsenaars  S.N.M.,
Newton-equivalent Hamiltonians for the harmonic oscillator,
{\it Ann. Physics} {\bf 293} (2001), 92--109.


\bibitem{turb2}
Turbiner  A.V.,
 Quantum mechanics: problems intermediate between exactly solvable and completely unsolvable,
 {\it Soviet Phys. JETP} {\bf 67} (1988), 230--236.\\
 Gonz\'arez-L\'opez  A., Kamran N., Olver  P.,
 Normalizability of one-dimensional quasi-exactly solvable Schr\"odinger operators,
{\it Comm. Math. Phys.} {\bf 153} (1993), 117--146.

\bibitem{turb3}
Smirnov   Y., Turbiner  A.,
Lie algebraic discretization of dif\/ferential equations,
{\it Modern Phys. Lett. A} {\bf 10} (1995), 1795--1802,
\href{http://arxiv.org/abs/funct-an/9501001}{funct-an/9501001}.\\
Chrissomalakos  C., Turbiner  A.,
Canonical commutation relation preserving maps,
{\it J. Phys.~A: Math. Gen.} {\bf 34} (2001), 10475--10485,
\href{http://arxiv.org/abs/math-ph/0104004}{math-ph/0104004}.

\bibitem{os14}
Odake  S., Sasaki  R.,
Unif\/ied theory of exactly and quasi-exactly solvable `discrete' quantum mechanics. I.~Formalism,
\href{http://arxiv.org/abs/0903.2604}{arXiv:0903.2604}.

\end{thebibliography}
\end{document}